\documentclass{aa}  

\usepackage{graphicx}
\usepackage{txfonts}
\begin{document}

   \title{A dense, solar metallicity ISM in the $z$=4.2 dusty star-forming galaxy SPT~0418-47}
\titlerunning{A dense, solar metallicity ISM in SPT~0418-47 at $z$=4.2}
   \author{Carlos De Breuck\inst{1}
          \and
          Axel Wei\ss \inst{2}
          \and
          Matthieu B\'ethermin \inst{3}
          \and
          Daniel Cunningham \inst{4,5}
          \and
          Yordanka Apostolovski \inst{6,7}
          \and
          Manuel Aravena \inst{8}
          \and
          Melanie Archipley \inst{9}
          \and
          Scott Chapman \inst{4,5,10,11}
          \and
          Chian-Chou Chen \inst{1}
          \and
          Jianyang Fu \inst{9}
          \and
          Sreevani Jarugula \inst{9}
          \and
          Matt Malkan \inst{12}
          \and
          Amelia C. Mangian \inst{9}
          \and
          Kedar A. Phadke \inst{9}
          \and
          Cassie A. Reuter \inst{9}
          \and
          Gordon Stacey \inst{13}
          \and
          Maria Strandet \inst{2}
          \and
          Joaquin Vieira \inst{9}
          \and
          Amit Vishwas \inst{13}
   }

   \institute{European Southern Observatory,
              Karl Schwarzschild Stra\ss e 2, 85748 Garching, Germany\\
              \email{cdebreuc@eso.org}
          \and
          Max-Planck-Institut für Radioastronomie, Auf dem Hügel 69, D-53121 Bonn, Germany
          \and
          Aix Marseille University, CNRS, LAM, Laboratoire d'Astrophysique de Marseille, Marseille, France
          \and
          Department of Astronomy and Physics, Saint Mary's University, Halifax, NS, B3H 3C3, Canada
          \and
          Department of Physics and Atmospheric Science, Dalhousie University, Halifax, NS, B3H 4R2, Canada
          \and
          Departamento de Ciencias Fisicas, Universidad Andres Bello, Fernandez Concha 700, Las Condes, Santiago, Chile
          \and
          Millennium Institute of Astrophysics (MAS), Nuncio Monseñor Sotero Sanz 100, Providencia, Santiago, Chile
          \and
          N\'ucleo de Astronom\'ia, Facultad de Ingenier\'ia, Universidad Diego Portales, Av. Ej\'ercito 441, Santiago, Chile
          \and
          Department of Astronomy, University of Illinois at Urbana-Champaign, 1002 West Green St., Urbana, IL 61801, USA
         \and
         National Research Council, Herzberg Astronomy and Astrophysics, Victoria, British Columbia, Canada
         \and
         Department of Physics and Astronomy, University of British Columbia, Vancouver, BC, V6T 1Z1, Canada
         \and
         Department of Physics and Astronomy, University of California, Los Angeles, CA 90095-1547, USA
         \and
         Department of Astronomy, Cornell University, Ithaca, NY 14853, USA}
   \date{Received 2019 June 24; accepted 2019 September 26}

 
  \abstract
   {We present a study of six far-infrared fine structure lines in the $z$=4.225 lensed dusty star-forming galaxy SPT~0418-47 to probe the physical conditions of its inter stellar medium (ISM). In particular, we report Atacama Pathfinder EXperiment (APEX) detections of the [OI]\,145$\mu$m and [OIII]\,88$\mu$m lines and Atacama Compact Array (ACA) detections of the [NII]\,122 and 205$\mu$m lines. The [OI]\,145$\mu$m / [CII]\,158$\mu$m line ratio is $\sim$5$\times$ higher compared to the average of local galaxies. We interpret this as evidence that the ISM is dominated by photo-dissociation regions with high gas densities. The line ratios, and in particular those of [OIII]\,88$\mu$m and [NII]\,122$\mu$m imply that the ISM in SPT~0418-47 is already chemically enriched close to solar metallicity. While the strong gravitational amplification was required to detect these lines with APEX, larger samples can be observed with the Atacama Large Millimeter/submillimeter Array (ALMA), and should allow observers to determine if the dense, solar metallicity ISM is common among these highly star-forming galaxies.}


   \maketitle
%

\section{Introduction}

The interaction between the interstellar medium (ISM) and stars is key to understanding galaxy formation and evolution. Stars form from dense, cool gas, and have a profound influence on their surrounding ISM by heating and ionizing the surrounding gas, which, in turn, inhibits further star formation. Stellar winds and supernovae enrich the ISM with newly formed metals, influencing the gas cooling rates, and hence the star-formation efficiency. A detailed knowledge of the physical parameters of the ISM, such as the metallicity, ionization state, density, and temperature is therefore essential to study galaxy evolution. The main diagnostics available over a wide range of redshifts are dust and stellar continuum emission as well as spectral emission and absorption lines.

The most frequently used methods to determine the star-formation rate (SFR) include direct far-ultraviolet (far-UV) emission from young stars, far-infrared (far-IR) dust continuum emission from dust heated by young stars, and recombination lines (e.g. H$\alpha$) ionized by the emission field of the young stellar population \citep[see][for a review]{kennicutt2012}. To determine the gas metallicity, rest-frame optical emission lines are traditionally the most frequently used tracers \citep[e.g.][]{maiolino2008}. The density, ionization parameter, and temperature of the gas can be derived from line ratios of fine structure lines (FSL) \citep[e.g.][]{villarmartin1997,debreuck2000}. One common disadvantage of these mostly optical tracers is that they can be severely affected by dust obscuration. Such problems can be avoided, or at least minimized by observing rest-frame far-IR FSL. While some galaxies such as Arp~220 are optically thick out to $\lambda_{\rm rest}$=240\,$\mu$m \citep{rangwala2011}, the extinction corrections to far-IR FSL are at most $\sim$50\% for the shortest wavelength lines \citep{uzgil2016}. Moreover, these far-IR FSL are often brighter than their optical counterparts \citep[e.g.][]{palay2012,moriwaki2018}; the brightest FSL of carbon, oxygen, and nitrogen are the major coolants of major phases in the ISM, and can each carry up to $\sim$1\% of the total far-IR luminosity.

The far-IR FSL originate both from photo-dissociation regions (PDR) and from HII regions, depending if their ionization potentials are lower or higher than that of hydrogen (13.6\,eV), respectively. All atomic FSL ([OI]\,63,145\,$\mu$m and [CI]~370,609\,$\mu$m) mainly originate from the neutral PDR. But also the ionic [CII]\,158\,$\mu$m line has an ionization potential of 11.26\,eV, and therefore originates from both the PDR and HII regions. Any diagnostic line ratios including such multiple-origin lines should therefore invoke a self-consistent model that treats both PDR and HII regions in pressure equilibrium \citep[e.g.][]{abel2005,nagao2012,cormier2019}. The relative contributions to composite lines such as [CII] can also be determined observationally by comparing the line flux with that from another FSL with a similar critical density that originates fully in the HII region. One such line is [NII]\,205\,$\mu$m, which has $n_{\rm crit}^{e^-}$=48\,cm$^{-3}$, compared to $n_{\rm crit}^{e^-}$=50\,cm$^{-3}$ for [CII]\footnote{This assumes the [CII]\,158\,$\mu$m is optically thin; in the SPT sample, the line appears to have moderate optical depth \citep{gullberg2015}. For optically thick [CII]\,158\,$\mu$m, the critical densities are lower by an order of magnitude, see Fig.~3 of \citet{lagache2018}.} at the same temperature \citep[assuming electrons are the main collision partners, see][]{fernandez-ontiveros2016}. This technique has been used on individual galaxies at high redshift \citep[e.g.][]{decarli2014}, as well as on samples of nearby galaxies \citep{kamenetzky2014,diaz-santos2017,croxall2017}, yielding [CII] PDR fractions of $\sim$75\%. The latter authors did find a clear dependence on metallicity in this PDR/HII fraction. If the fraction can be found by other means, this dependence can also be turned around to determine the gas metallicity independently of classical optical diagnostics \citep[e.g.][]{nagao2011,nagao2012,debreuck2014,bethermin2016,pereira-santaella2017}. An easier solution is to consider only HII-dominated FSL to determine the physical parameters. For example the [NII]\,122\,$\mu$m/[NII]\,205\,$\mu$m and [OIII]\,52\,$\mu$m/[OIII]\,88\,$\mu$m ratios are powerful density tracers in the range 1--10$^3$ and 10--10$^{4.5}$\,cm$^{-3}$, respectively \citep{fernandez-ontiveros2016}. The most promising metallicity tracers are the [OIII]\,88\,$\mu$m/[NII]\,122\,$\mu$m or [OIII]\,88\,$\mu$m/[NII]\,205\,$\mu$m ratios \citep{pereira-santaella2017}. Using {\it Herschel} Spectral and Photometric Imaging Receiver (SPIRE) spectroscopy of three gravitationally lensed dusty star-forming galaxies (DSFG), \citet{rigopoulou2018} use the [OIII]\,88\,$\mu$m/[NII]\,122\,$\mu$m ratio to argue that these DSFGs must already have close to solar metallicities. \citet{tadaki2019} confirm a similar result on an unlensed DSFG observed with the Atacama Large Millimeter/submillimeter Array (ALMA).

The applicability of mid-IR and far-IR FSL as diagnostic tools of the ISM has received a boost thanks to the publication of samples of nearby galaxies observed with the infrared space observatory ({\it ISO}) and {\it Herschel} \citep[e.g.][]{brauher2008,farrah2013,sargsyan2014,kamenetzky2014,cormier2015,cigan2016,herrera-camus2016,fernandez-ontiveros2016,zhao2016,diaz-santos2017,zhang2018}. At high redshift ($z$$\gtrsim$1), the far-IR FSL conveniently shift into the (sub)millimetre atmospheric windows. The most popular line is clearly [CII]\,158\,$\mu$m, followed by the [CI]~370,609\,$\mu$m lines \citep[e.g.][]{walter2011,bothwell2017}. Both single dish submillimetre telescopes and interferometers have also detected the [NII]\,122\,$\mu$m and 205\,$\mu$m lines at high redshift \citep{ferkinhoff2011,ferkinhoff2015,combes2012,nagao2012,decarli2012,decarli2014,bethermin2016,pavesi2016,lu2017,tadaki2019,novak2019}. After a slow start, the [OIII]\,88\,$\mu$m line is quickly becoming a popular line to confirm redshifts of galaxies in the epoch of reionization ($z\gtrsim 6$), where it shifts into the submillimetre atmospheric windows below 500\,GHz \citep{ferkinhoff2010,inoue2016,carniani2017,marrone2018,vishwas18,hashimoto2018,walter2018,tamura2019,tadaki2019,hashimoto2019,novak2019}.
Deep {\it Herschel}/SPIRE spectroscopy has also revealed a number of FSL detections, either in individual objects \citep{valtchanov2011,coppin2012,george2013,uzgil2016,rigopoulou2018,zhang2018}, or in stacked spectra \citep{wardlow2017,wilson2017,zhang2018}. These include the only detections of the [OI]\,63\,$\mu$m line at high redshift reported thus far. This [OI]$^3P_2 - ^3P_1$ line is arguably the best tracer for the star-forming gas, as it traces the very dense ($n_{\rm crit}^H$=5$\times$10$^5$\,cm$^{-3}$) neutral gas (one caveat being the frequeny presence of self-absorption observed in local ultra-luminous IR galaxies, \citet{rosenberg2015}). Like the [CII] line, the [OI]\,63\,$\mu$m line also shows a 'deficit' in the most luminous FIR sources, though with a higher scatter \citep{gracia-carpio2011,cormier2015,diaz-santos2017}. Surprisingly, this bright FSL has not been frequently observed with ALMA, probably because it is only observable in the highest frequency bands. At least as surprising is that the fainter, but more accessible [OI]$^3P_1 - ^3P_0$ transition at $\lambda_{\rm rest}$=145\,$\mu$m ($n_{\rm crit}^H$=9.5$\times$10$^4$\,cm$^{-3}$) has thus far not been detected at high redshifts (only \citet{novak2019} report a tentative detection in a $z$=7.5 quasar). Also in nearby galaxies, this [OI]\,145$\mu$m line has not been observed very frequently as in most cases, it is fainter than the nearby [CII]\,158$\mu$m line. After initial detections with {\it ISO} \citep{malhotra2001,brauher2008}, {\it Herschel} has now detected [OI]\,145$\mu$m in significant samples of nearby galaxies \citep{spinoglio2015,cormier2015,fernandez-ontiveros2016,herrera-camus2018a}, and recently in a $z$=6.5 lensed quasar \citep{yang2019}.

In this paper, we report the second such [OI]\,145\,$\mu$m detection in SPT-S~J041839-4751.8 (hereinafter SPT~0418-47) at $z$=4.2248 \citep{weiss2013}. With a lensing magnification $\mu$=32.7$\pm$2.7 \citep{spilker2016}, SPT~0418-47 is one of the most strongly lensed DSFGs from our sample selected from the South Pole Telescope, making its FSL accessible to the Atacama Pathfinder EXperiment (APEX). We also report the detection of the [OIII]\,88\,$\mu$m line, and use the five detected FSL in SPT~0418-47 to constrain the metallicity in this DSFG. Throughout this paper, we assume a $\Lambda$CDM cosmology with $H_0$=67.8\,s$^{-1}$Mpc$^{-1}$, $\Omega_m$=0.308, and $\Omega_{\Lambda}$=0.692 \citep{planck2016}. At $z$=4.2248, this corresponds to a luminosity distance $D_L$=39.150\,Gpc and a scale of 7\,kpc/\arcsec.
\begin{figure*}[ht]
\centering
\begin{tabular}{ccccc}        
\includegraphics[width=3.2cm]{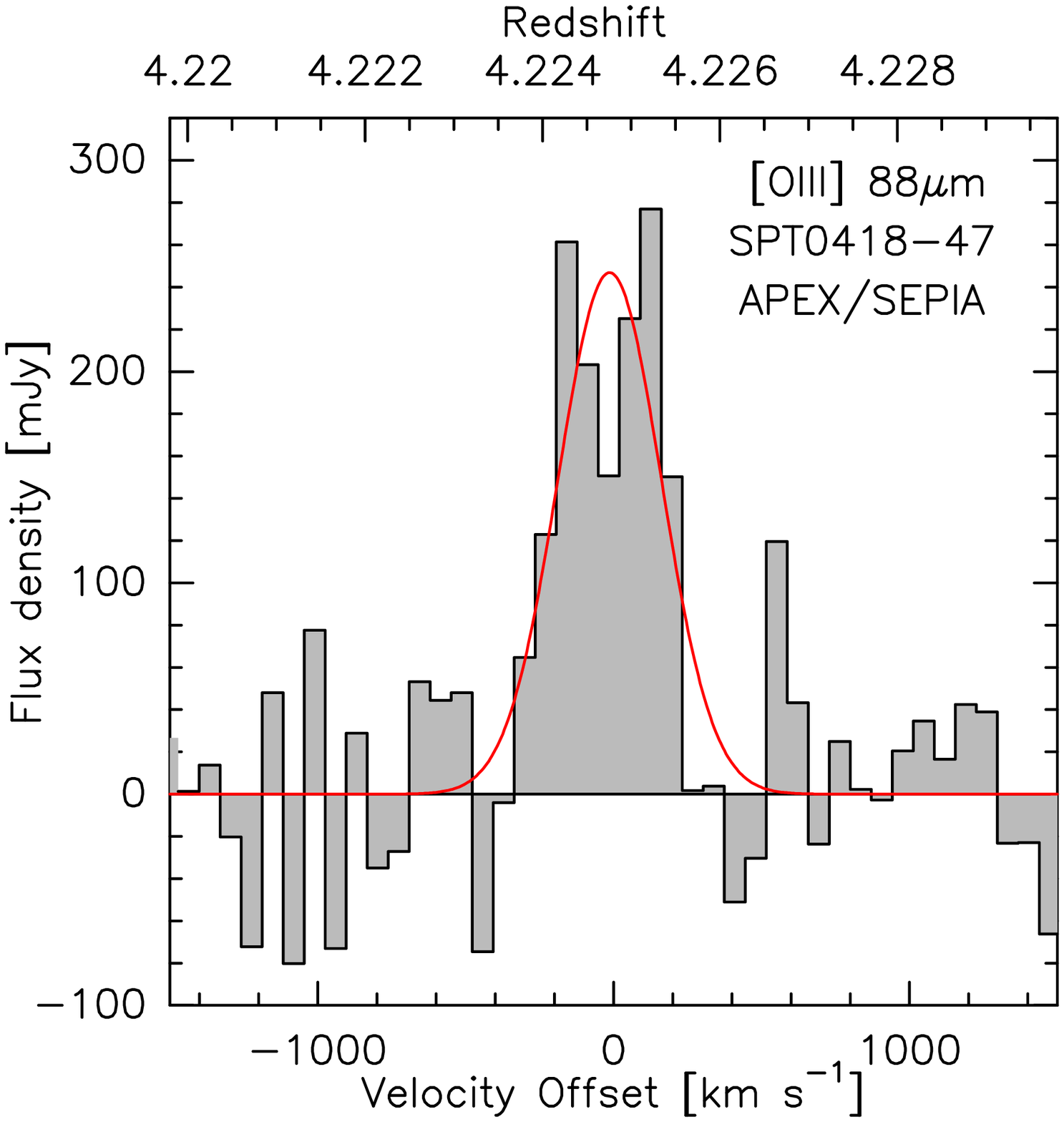} &
\includegraphics[width=3.1cm]{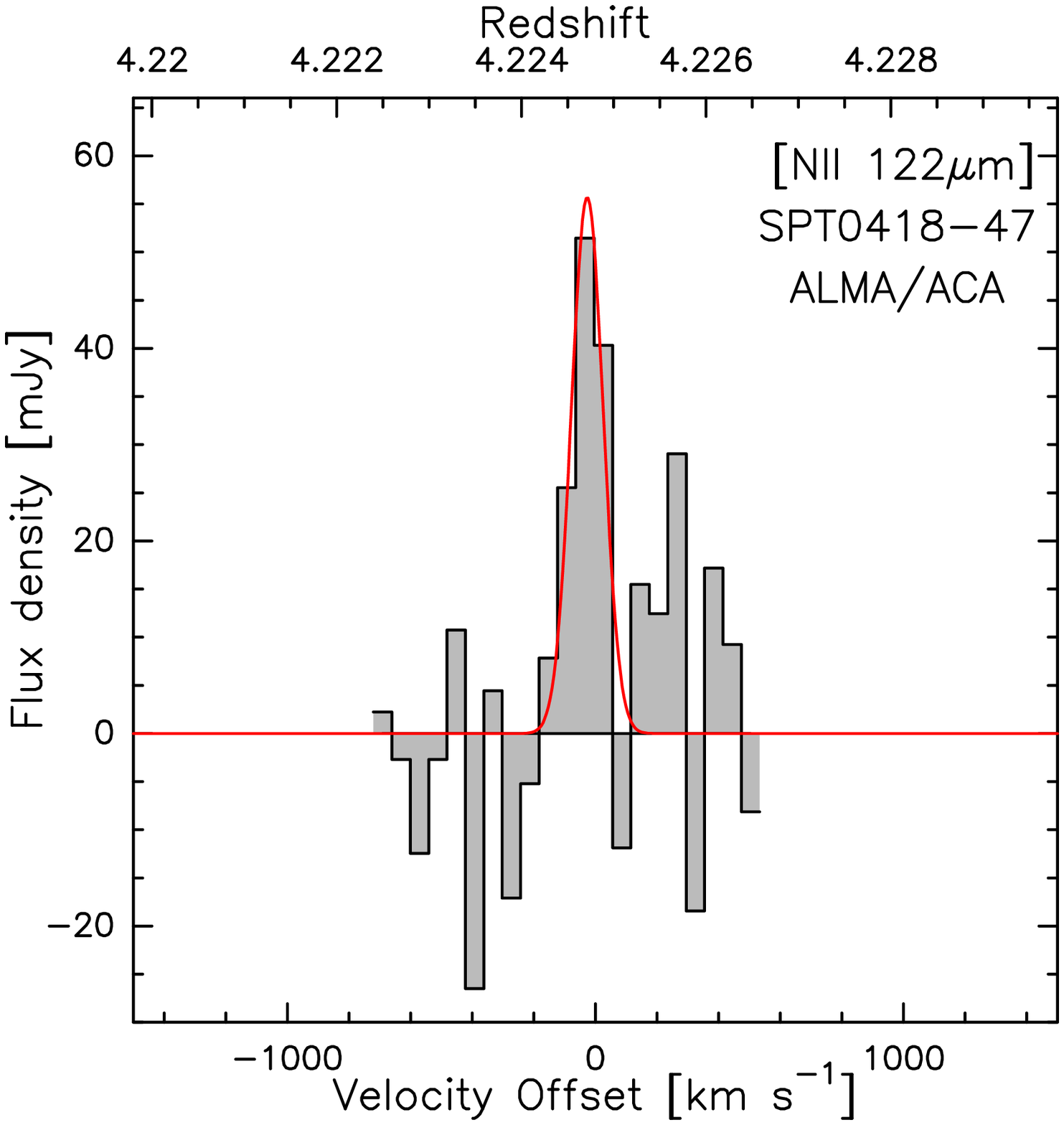} &
\includegraphics[width=3.5cm]{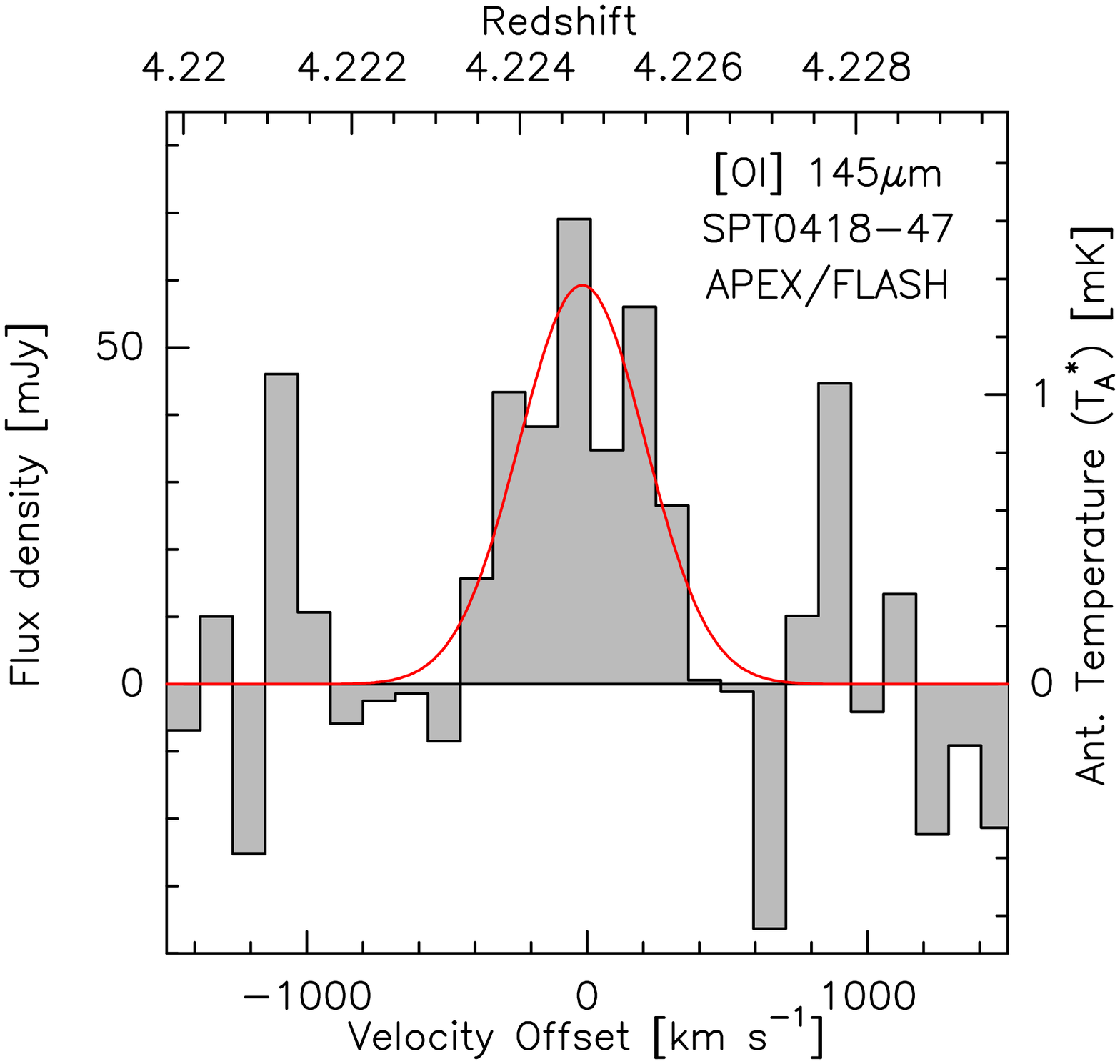} &
\includegraphics[width=3.5cm]{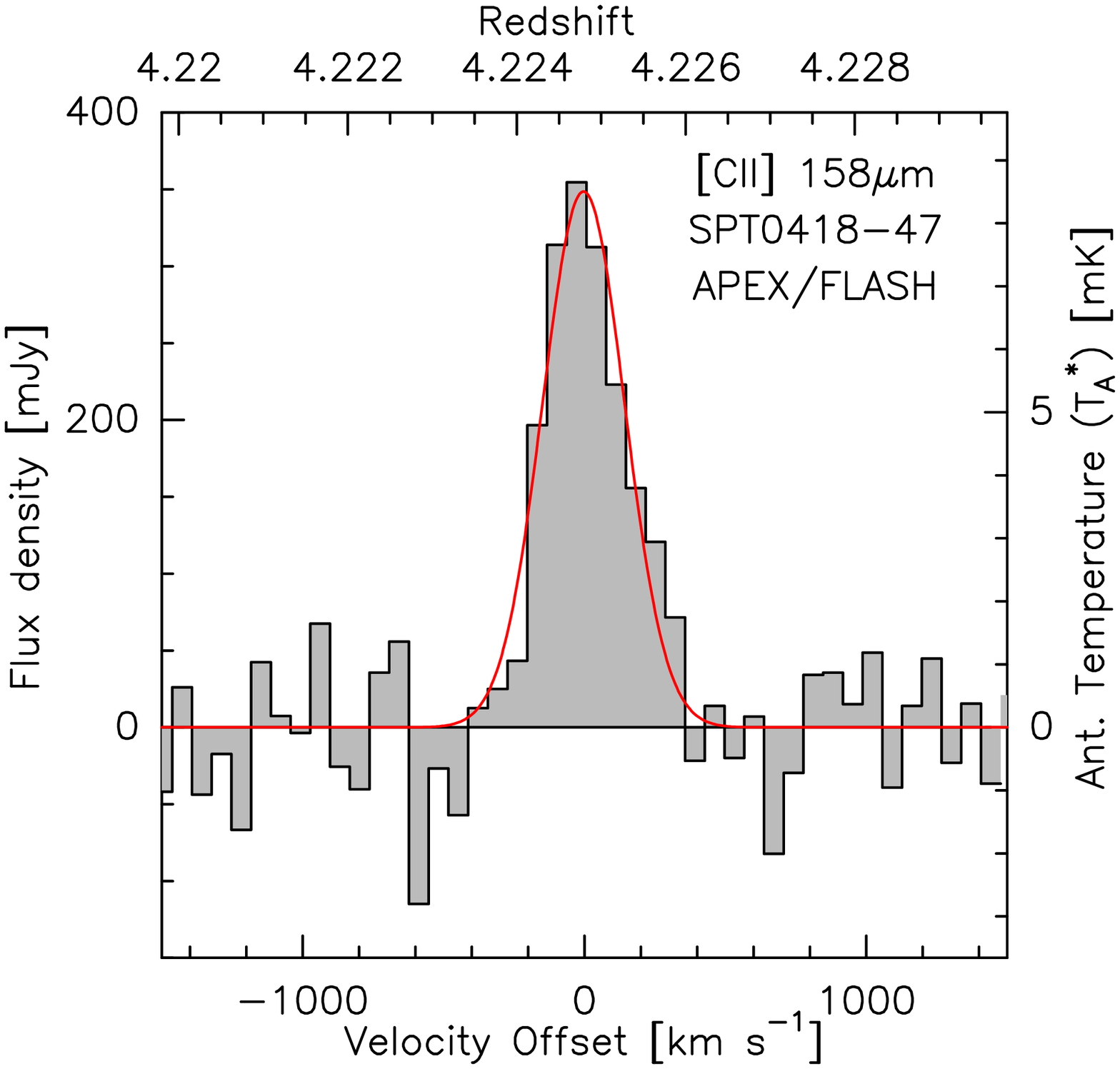} &
\includegraphics[width=3.1cm]{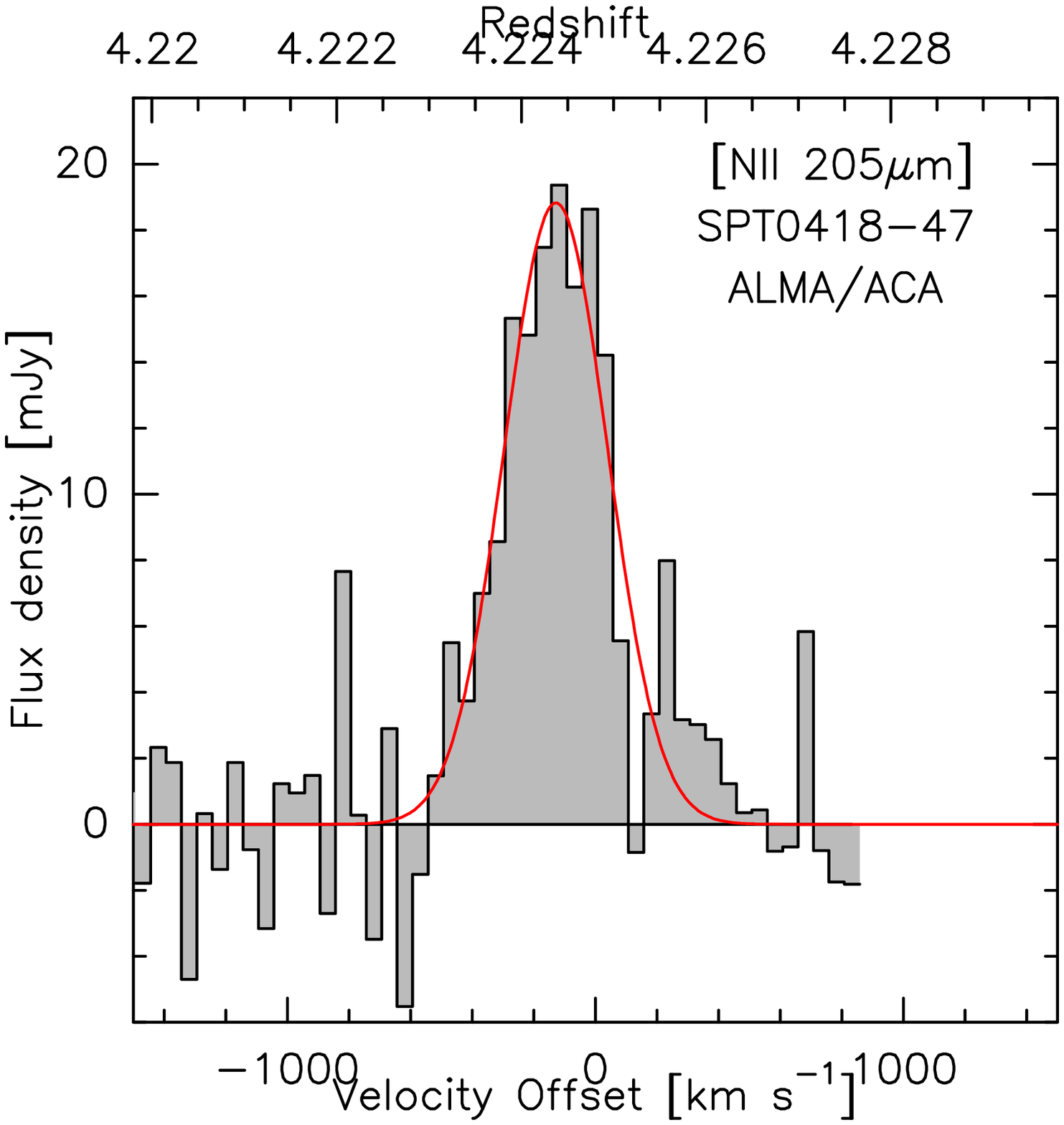} \\
\end{tabular}
\caption{Spectra of [OIII]\,88\,$\mu$m, [NII]\,122\,$\mu$m, [OI]\,145\,$\mu$m, [CII]\,158\,$\mu$m \citep{gullberg2015}, and [NII]\,205\,$\mu$m \citep{cunningham2019}, fine structure lines in SPT~0418-47. The channel widths are 71, 60, 116, 70 and 50\,km/s, respectively. All velocities are given relative to $z$=4.2248.}
\label{Spectra}%
\end{figure*}
%
%
%
\begin{table*}[ht]
\caption{Observational parameters of the fine structure lines in SPT~0418-47 (no gravitational amplification corrections applied)} 
\label{obsparams}      
\begin{center}                          
\begin{small}
\begin{tabular}{l c c c c c c}        
\hline\hline                 
Line & [OIII]\,88\,$\mu$m & [NII]\,122\,$\mu$m & [OI]\,145\,$\mu$m & [CII]\,158\,$\mu$m & [NII]\,205\,$\mu$m & [CI]~609\,$\mu$m \\    
\hline                        
Redshift  & 4.2246$\pm$0.0004 & 4.2242$\pm$0.0003 & 4.225$\pm$0.001 & 4.2248$\pm$0.0002 & 4.2248$\pm$0.0005 & 4.225$\pm$0.004 \\
$\Delta V$ [km/s] & 400$\pm$40 & 130$\pm$50 & 530$\pm$110 & 342$\pm$31 & 400$\pm$40 & 310$\pm$50 \\
$F_{\rm line}$ [Jy km/s] & 108$\pm$11 & 7.1$\pm$2.0 & 33$\pm$7 & 127$\pm$10 & 7.9$\pm$0.6 & 2.5$\pm$0.6 \\
$L_{\rm Line}$ [10$^{10}$ L$_{\odot}$] & 11.1$\pm$1.2 & 0.53$\pm$0.15 & 2.1$\pm$0.5 & 7.4$\pm$0.6 & 0.35$\pm$0.03 & 0.037$\pm$0.009 \\
Reference$^a$ & [1] & [2] & [1] & [3] & [4] & [5] \\
\hline                                   
\end{tabular}
\end{small}
\end{center}
$^a$ [1] this paper; [2] Cunningham et al., in prep.; [3] \citet{gullberg2015}; [4] \citet{cunningham2019} [5] \citet{bothwell2017}.
\end{table*}
%
%
%
\begin{table*}[ht]
\caption{Continuum photometry of SPT~0418-47, taken from \citet{strandet2018} and Reuter et al, in prep. Fluxes are given in mJy.} 
\label{photometry}      
\centering                          
\begin{tabular}{ccccccccc}        
\hline\hline                 
$S_{3.0mm}$ & $S_{2.0mm}$ & $S_{1.4mm}$ & $S_{870\mu m}$ & $S_{500\mu m}$ & $S_{350\mu m}$ & $S_{250\mu m}$ & $S_{160\mu m}$ & $S_{100\mu m}$ \\
\hline                                   
0.79$\pm$0.13 & 7.5$\pm$1.2 & 35.9$\pm$6.4 & 108$\pm$10 & 175$\pm$7 & 166$\pm$6 & 114$\pm$6 & 45$\pm$8 & $<$6.9 \\
\hline                                   
\end{tabular}
\end{table*}
%

\section{Observations}
\subsection{New data from APEX}
We observed the [OI]\,145\,$\mu$m ($\nu_{\rm rest}$=2060.07\,GHz) using the First-Light APEX Submillimeter Heterodyne instrument \citep[FLASH;][]{klein2014} on the APEX telescope \citep{guesten2006}. The observations were taken under the Max Planck project 093.F-9512, from 2014 May 14 to 20. The total on-source integration time was 7.3 hours, with precipitable water vapour (PWV) in the range 0.4--0.8\,mm. The receiver was tuned to 394.286\,GHz in the lower sideband, and used the wobbler with an amplitude of 30\arcsec\ and frequency of 1.5\,Hz. We reduced the data using the standard procedures in the Continuum and Line Analysis Single-dish Software \citep[CLASS;][]{pety2005}, and used an antenna gain of Jy/K=43. We estimate the overall calibration uncertainty at 15\%.

To observe the [OIII]\,88\,$\mu$m line ($\nu_{\rm rest}$=3393.01\,GHz), we used the Swedish-ESO PI Instrument for APEX \citep[SEPIA;][]{belitsky2018}. The observations were taken under the ESO project E-098.A-0513, on 2016 August 10 and 2017 April 26 with the first double sideband (DSB) incarnation of the receiver and the 4\,GHz wide XFFTS backend, and under the Max Planck account M-0101.F-9522 on 2019 June 8 with the final sideband-separed (2SB) version of the receiver and the 8\,GHz wide dFFTS4G backend. The total on-source integration time was 6.25 hours, with precipitable water vapour (PWV) in the range 0.25--0.5\,mm. The receiver was tuned to 649.403\,GHz in the lower sideband, and used the wobbler with an amplitude of 40\arcsec\ and frequency of 0.5\,Hz in 2016/2017 and 30\arcsec with 1.5\,Hz in 2019. We reduced the data using the standard procedures in CLASS, using a first order fit for baseline removal. We used antenna gains of Jy/K=120 for the DSB receiver and Jy/K=70 for the 2SB receiver, based on carefully pointed and focused observations of Uranus close in time to the science target. We re-observed the 8.3$\sigma$ DSB line with the 2SB receiver because after the DSB observations, a construction problem with the SEPIA band 9 selection mirror \citep[NMF3 in Fig. 1 and 3 of][]{belitsky2018} was identified (and soon thereafter corrected). As a result, there were time-variable optical losses, which are difficult to characterize for long integrations like ours. This effect cannot be separated from other changes between 2016 and 2019, including a move of SEPIA to the central position in the Nasmyth A cabin using a new set of mirrors, the change of receiver from DSB to 2SB, the change of the telescope subreflector, and the replacement of the main telescope mirror surface panels to improve the overall telescope efficiency. However, thanks to the frequent planet calibration observations, we were able to combine these changes into the 70\% improvement in the antenna gain, leading to consistent [OIII]\,88\,$\mu$m line detections of 108$\pm$13 and 82$\pm$16\,Jy\,km\,s$^{-1}$ for the DSB and 2SB receivers, respectively. For our analysis, we use the weigthed average flux of 105$\pm$11\,Jy\,km\,s$^{-1}$.

\subsection{Additional data}
In addition to the two FSL reported above, we also include three other lines in our analysis. \citet{gullberg2015} report a bright [CII]\,158\,$\mu$m line detected with APEX/FLASH, reproduced next to the [OIII] and [OI] lines in Fig.~\ref{Spectra}. \citet{bothwell2017} report the detection of the [CI]~609$\mu$m line from ALMA Cycle~3 spectroscopy. The [NII]\,205$\mu$m line was detected as part of an Atacama Compact Array (ACA) survey of 41 SPT sources \citep{cunningham2019}. The [NII]\,122$\mu$m line was detected in a similar continuation of this ACA project (Cunningham et al., in prep.). Table~\ref{obsparams} lists the main parameters from these observations. 

The dust continuum SED has been sampled by seven photometric points (Table~\ref{photometry}), yielding a dust temperature $T_d$=45$\pm$2\,K \citep{strandet2016} and intrinsic FIR (8--1000$\mu$m) luminosity $L_{\rm FIR}$=(2.8$\pm$0.3)$\times10^{12}$L$_{\odot}$ \citep{bothwell2017}. The lensing amplification factor $\mu$=32.7$\pm$2.7 was derived by \citet{spilker2016} based on 0$\farcs$5 resolution ALMA imaging. The lens model also shows that the lensed galaxy is very compact with an effective radius of 0\farcs092$\pm$0.008 ($\sim$0.6\,kpc).
\begin{figure}[ht]
\centering
\includegraphics[width=9cm]{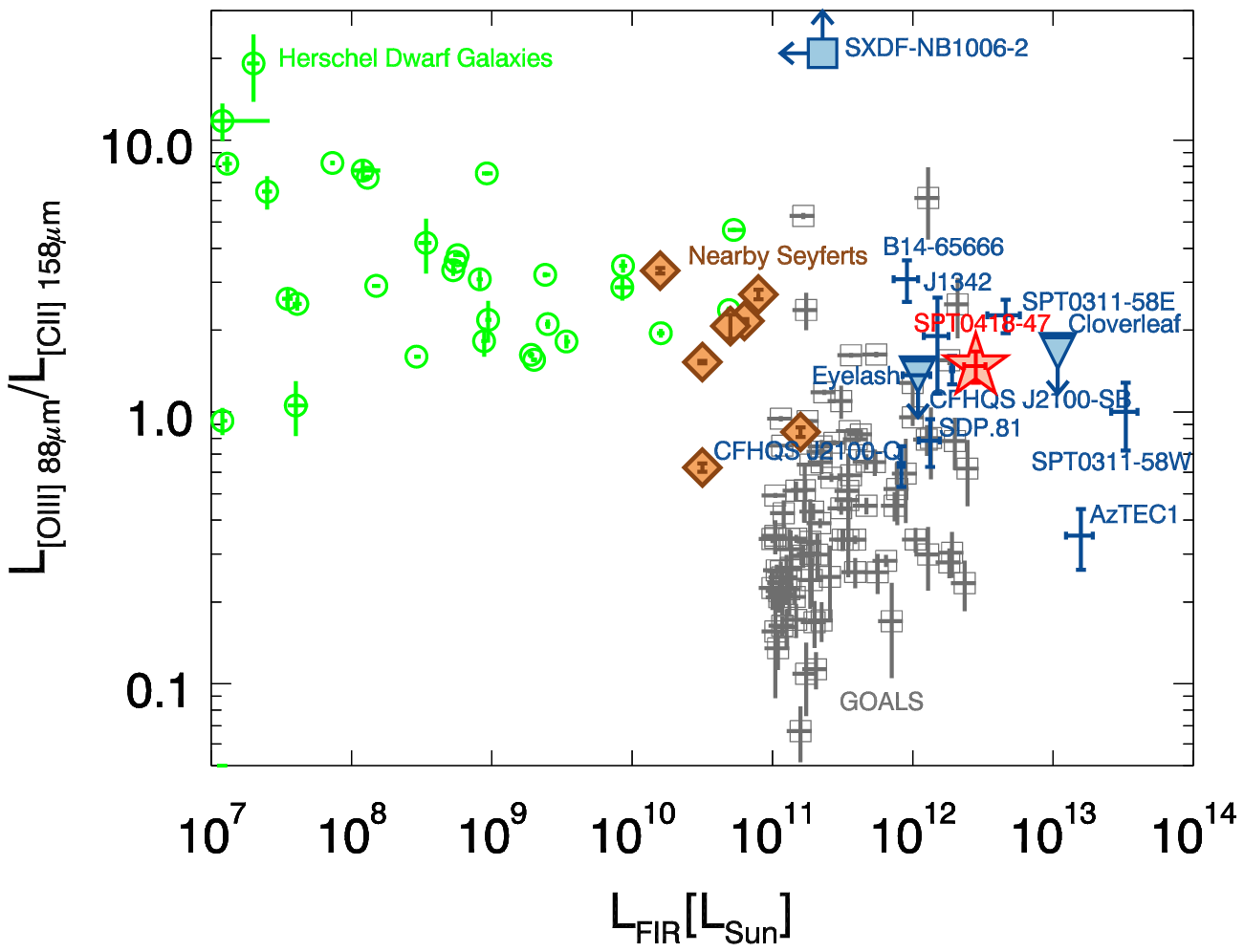} \\
\caption{Ratio of the [OIII]\,88\,$\mu$m and [CII]\,158\,$\mu$m line luminosities as a function of (intrinsic) FIR luminosity. The comparison low-$z$ samples are the {\it Herschel} Dwarf Galaxies \citep[green circles;][]{cormier2015}, nearby Seyferts \citep[orange diamonds;][]{spinoglio2015}, and the GOALS sample \citep[grey squares, assuming a 20\% uncertainty in $L_{\rm FIR}$;][]{sanders2003,diaz-santos2017}. The high-$z$ sources marked in blue are SDP.81 \citep{valtchanov2011}, the Eyelash \citep{george2013}, the Cloverleaf \citep{uzgil2016}, SXDF-NB1006-2 \citep{inoue2016}, SPT~0311-58 \citep{marrone2018}, CFHQS~J2100 \citep{walter2018}, COSMOS-AzTEC1 \citep{tadaki2019}, B14-65666 \citep{hashimoto2019}, and J1342+0928 \citep{novak2019}. }
\label{CII-OIII_vs_LIR}%
\end{figure}

\section{Results}
Fig.~\ref{Spectra} shows our detections of both the [OIII]\,88 \,$\mu$m and [OI]\,145\,$\mu$m lines at the $\sim$5$\sigma$ level. We fitted an unconstrained single Gaussian to the observed profiles and list the redshift, velocity width $\Delta V$ and integrated line luminosity in Table~\ref{obsparams}. For consistency, we have used the same fitting procedure for all lines shown in Fig.~\ref{Spectra}. We note that the [NII]\,122\,$\mu$m line width is less than half that of the other lines, but this may be due to the low signal to noise ratio (S/N) of our detection. Integrating over $-$300 to +350km\,s$^{-1}$, that is the full velocity width of the bright [CII]\,158\,$\mu$m line yields an integrated [NII]\,122\,$\mu$m line flux of 7.1$\pm$2.5\,Jy\,km\,s$^{-1}$. This is fully consistent with our unconstrained Gaussian fit.  Deeper data are required to obtain a more reliable line flux, but the uncertainties are not to a level that they would change the conclusions in this paper.

The line luminosities are calculated following \citet{solomon1997}:
$$L_{\rm line}=1.04\times10^{-3}S_{\rm line}\Delta V\nu_{\rm rest}(1+z)^{-1}D^2_{\rm L},$$ where $S_{\rm line}\Delta V$ is the velocity-integrated line flux in Jy\,km/s and $\nu_{\rm rest}$ is the rest frequency in GHz. The redshifts and velocity widths are consistent with each other within the uncertainties for all FSL, with the exception of the low S/N [NII]\,122$\mu$m line. The spectral profiles of the five well-detected FSL are also consistent with the CO(2-1) and CO(4-3) lines, suggesting that differential lensing is not important in this source \citep{gullberg2015,aravena2016}. Moreover there is no evidence for any companion galaxies in the lens model of SPT~0418-47 \citep{spilker2016}, so we can safely assume that the integrated line fluxes are tracing a single galaxy. 

We first compare our new FSL detections with other high-$z$ galaxies. The only FSL that have been studied in significant samples of high-$z$ galaxies are [CII] and [CI] \citep[][]{stacey2010,walter2011,alaghband-zadeh2013,gullberg2015,bothwell2017,lagache2018}. In both of these lines, the luminosity of SPT~0418-47 falls near the average of the SPT DSFGs. A first survey of the [NII]\,205$\mu$m lines has been observed in 41 galaxies from our SPT sample \citep{cunningham2019}; SPT~0418-47 has been observed as part of this survey.

Detections of the [OIII]\,88\,$\mu$m line have now been reported in a dozen other high-$z$ objects. Fig.~\ref{CII-OIII_vs_LIR} compares the ratio of those [OIII]\,88\,$\mu$m lines with [CII]\,158\,$\mu$m as a function of $L_{\rm FIR}$. Compared to eleven other high-$z$ DSFGs, SPT~0418-47 has one of the brightest relative [OIII]\,88\,$\mu$m emission, emitting about twice as much power as the [CII]\,158\,$\mu$m line. Only the $z$=7.2 Lyman-$\alpha$ emitter SXDF-NB1006-2 has a ratio that is at least an order of magnitude higher, which has been argued as evidence for a very low metallicity \citep{inoue2016}. Interestingly, SPT~0418-47 falls close to the starburst galaxy rather than the quasar in the double system CFHQS~J2100 \citep{walter2018}. The only low-$z$ sample that reaches such similarly high $L_{\rm FIR}$ is the Great Observatories All-sky LIRG Survey \citep[GOALS;][]{sanders2003,diaz-santos2017}. Compared to this sample, SPT~0418-47 has a high [OIII]\,88\,$\mu$m/ [CII]\,158\,$\mu$m ratio, but there appears to be an increasing trend (and scatter) towards the highest $L_{\rm FIR}$. SPT~0418-47 appears to be more consistent with both the nearby dwarf galaxies \citep{cormier2015} and Seyferts \citep{spinoglio2015}. Increasing the number statistics is clearly required to perform a better comparison, but our bright [OIII]\,88\,$\mu$m detection suggests that detecting this line may be easier than suggested by the previously reported upper limits. 

\begin{figure*}[ht]
\centering
\includegraphics[width=18cm]{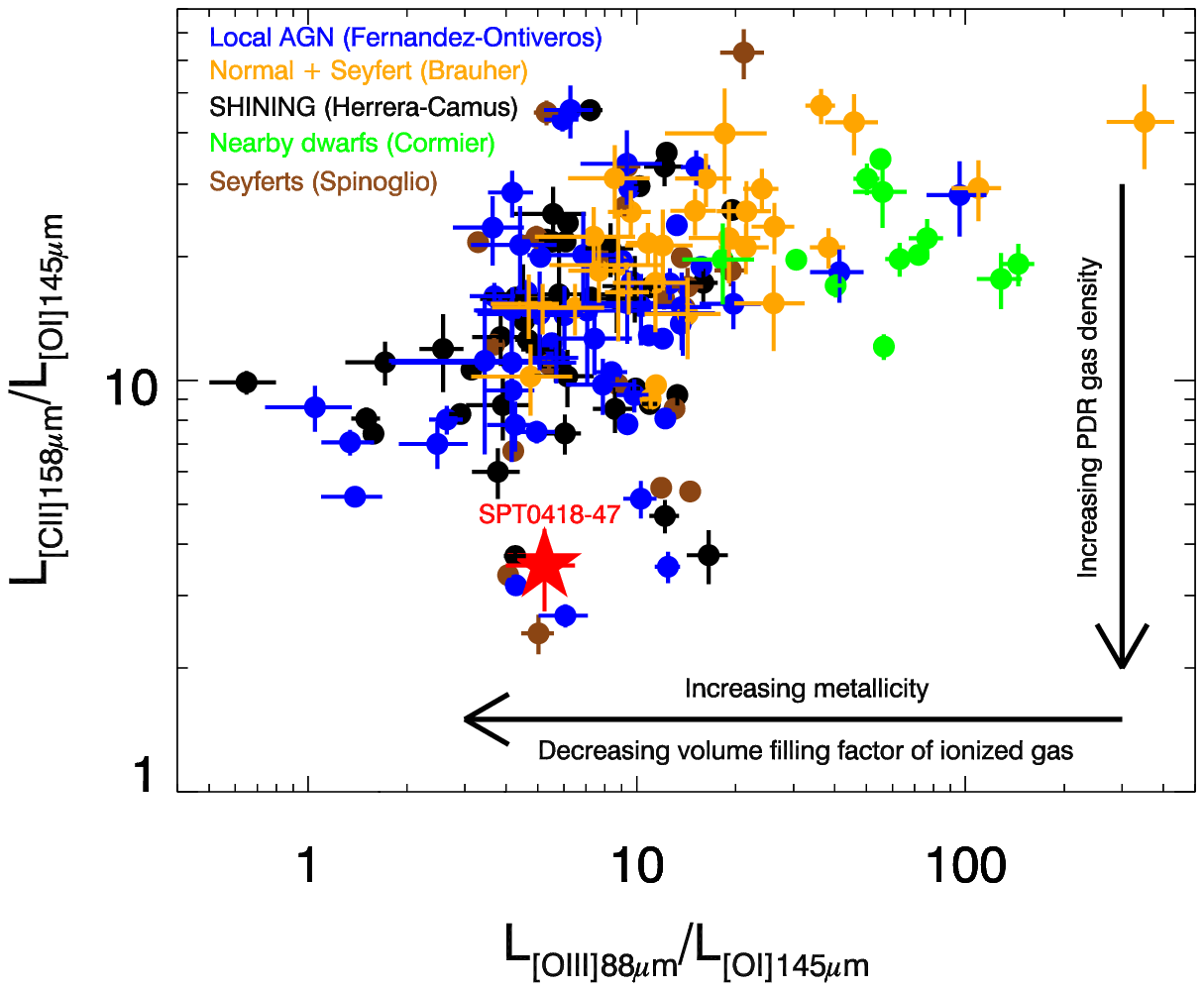} \\
\caption{Line luminosity ratio of [CII]\,158\,$\mu$m and [OI]\,145\,$\mu$m against [OIII]\,88 \,$\mu$m and [OI]\,145\,$\mu$m. The low-$z$ samples are drawn from \citet{fernandez-ontiveros2016}, \citet{brauher2008}, \citet{herrera-camus2018a}, \citet{cormier2015} and \citet{spinoglio2015}. We note that relative to [OI]\,145\,$\mu$m, SPT~0418-47 (red star) has an [OIII]\,88\,$\mu$m luminosity that is rather consistent with normal (black, brown and blue) rather than dwarf galaxies (green), but a weaker [CII]\,158\,$\mu$m. Arrows indicate which physical processes dominate these line ratios (see text).}
\label{CII-OIII-OI-ratio}%
\end{figure*}

Thus far, only one other [OI]\,145$\mu$m detections has been reported at $z$$>$1 \citep{yang2019}. Although there are a few detections of the 63\,$\mu$m line especially in stacked {\it Herschel} spectra \citep{uzgil2016,wardlow2017,wilson2017,zhang2018}, this is not the case in the fainter 145\,$\mu$m line. To check if our detection is expected, we scale from the [OIII]\,88\,$\mu$m and [CII]\,158\,$\mu$m detections in the stacked spectrum of DSFGs \citep{zhang2018}, which predicts a 63\,$\mu$m luminosity of 1.6--5.5\,$\times$10$^{10}$L$_{\odot}$. To convert this to an [OI]\,145$\mu$m luminosity, we use the observed 145\,$\mu$m to 63\,$\mu$m [OI] line luminosity ratio between 2\% and 35\% \citep{brauher2008,vasta2010}, which yields a predicted range of 0.02--1.9$\times$10$^{10}$L$_{\odot}$. Our detection at the 2.1$\times$10$^{10}$L$_{\odot}$ level (Table~\ref{obsparams}) is at the upper bound of this range. However, we warn that the ratio of the two [OI] lines is subject to extinction and self-absorption effects \citep[e.g.][]{vasta2010}, and is therefore quite uncertain. 
We therefore also directly compare our observed [OI]\,145$\mu$m detection with low-$z$ samples.  Fig.~\ref{CII-OIII-OI-ratio} plots the [CII]\,158$\mu$m and [OIII]\,88\,$\mu$m both normalized by the [OI]\,145\,$\mu$m luminosity. While the [OIII]\,88\,$\mu$m/[OI]\,145\,$\mu$m ratio is consistent with most normal galaxies (except dwarfs), the [CII]\,158$\mu$m in SPT~0418-47 appears to be underluminous relative to the [OI]\,145\,$\mu$m. In the next section, we combine all FSL results to determine the physical conditions of the ISM in SPT~0418-47.

\section{Discussion}
\subsection{The nature of SPT0418-47}\label{subsection:nature}
Before discussing the physical parameters of its ISM, we first compare SPT0418-47 with the main sequence galaxy population at $z$$\sim$4. Unfortunately, a direct stellar mass determination is rather complex because the foreground lensing galaxy outshines the background DSFG \citep{vieira2013}. Using the CIGALE SED modelling described in \S\ref{subsection:AGNcontributions} and the deblending technique of \citet{ma2015}, we derive an intrinsic $M_*$=(9.5$\pm$3.0)$\times 10^9$M$_{\odot}$ (Phadke et al., in prep.). This value can be compared with the stellar masses 2--8$\times$10$^{10}$M$_{\odot}$ in four other SPT DSFGs derived using {\it HST} and {\it Spitzer} imaging \citep{ma2015}. Using the gas masses derived from low-J CO lines \citep{aravena2016} in these four DSFGs, this implies stellar to gas mass ratios in the range 0.3 to 1.0. The relatively low gas mass in SPT~0418-47 of 0.5$\times$10$^{10}$M$_{\odot}$ then suggests the stellar mass in the range 0.5--1.7$\times$10$^{10}$M$_{\odot}$, fully consistent with the SED-derived $M_*$ above. Detailed {\it JWST} imaging of SPT0418-47 scheduled as part of the Early Release Science observations will provide a much more reliable stellar mass determination. 

This rather low stellar mass contrasts with the high SFR=280$\pm$30\,M$_{\odot}$yr$^{-1}$ derived as 10$^{-10}$$\times$$L_{\rm FIR}$. This places SPT0418-47 well above the galaxy main sequence at $z$$\sim$4, which predicts a SFR$\sim$25\,M$_{\odot}$yr$^{-1}$ for a 10$^{10}$\,M$_{\odot}$ galaxy \citep{schreiber2017}.

\subsection{Contributions of PDR and HII regions}\label{subsection:PDR_HII}

As mentioned in the introduction, the FSL we observed originate from both the PDR and HII regions. It is important to keep in mind that we are only looking at galaxy-integrated line luminosities rather than ratios within a single Giant Molecular Cloud (GMC). What we are comparing are thus the luminosity-weighted combinations of all GMCs within these galaxies.

The line that has the most uncertain relative contribution of PDR and HII regions is [CII]\,158\,$\mu$m. To estimate the expected contribution of HII regions to the [CII]\,158\,$\mu$m line flux, we can use the [NII]\,205\,$\mu$m line, which has the same critical density and similar ionization range. Following \citet{diaz-santos2017}, we assume $L_{[CII]}^{ion}/L_{[NII]}$$\simeq$3.0$\pm$0.5, which covers the usual full range of HII region physical conditions. Comparing this to the observed ratio of 24.5$\pm$2.0 (Table~\ref{obsparams}), we derive an HII fraction of 12$\pm$2\%. While there is some remaining model uncertainty on this number, it seems clear that the [CII] line is dominated by PDRs. Given that this contribution is rather small, we do not correct for it in the following; moreover such a correction cannot be calculated for all comparison samples due to the absence of systematic [NII]\,205\,$\mu$m observations.

Another line ratio that can be used to determine the ratio of the filling factors of warm ionized and dense neutral gas is [OIII]\,88\,$\mu$m / [OI]\,63\,$\mu$m \citep{cormier2015,cormier2019,diaz-santos2017}. This ratio provides a measure that is completely independent of abundance, and only very weakly on optical depth. The ratio could depend on temperature, though it would in this case measure the variation in the ratio of temperatures in the PDR and HII phases between galaxies. \citet{diaz-santos2017} (their Fig.~7) do not find such a trend in the local luminious infrared galaxies from GOALS. The only galaxies that show clearly higher [OIII]\,88\,$\mu$m / [OI]\,63\,$\mu$m ratios are dwarf galaxies, which has been interpreted as indicating the presence of low density channels allowing the hard UV photons to reach the outer parts of galaxies \citep{cormier2015,cormier2019}. As a result, these low metallicity galaxies are likely to be filled with relatively low density (below $n_{\rm crit}^{e^-}(\rm [OIII]\,88\,\mu$m)$\simeq$500\,cm$^{-3}$) gas rather than PDRs. One uncertainty in this ratio remains the optical depth and self-absorption of the [OI]\,63\,$\mu$m line. We here use instead the [OI]\,145\,$\mu$m line, which is not subject to these uncertainties. Fig.~\ref{CII-OIII-OI-ratio} confirms that the dwarf galaxies (green points) are clearly offset from the other samples, with the normal galaxies and Seyferts of \citet{brauher2008} (orange points) lying in between. SPT~0418-47 instead lies at the lower end of the [OIII]\,88\,$\mu$m / [OI]\,145\,$\mu$m distribution, which implies that the galaxy-integrated line emission is dominated by PDR. This confirms our conclusion above from the [CII]\,158\,$\mu$m / [NII]\,205\,$\mu$m ratio.

\subsection{Gas density}\label{subsection:density}

To determine the gas density, one should ideally use the ratio of two FSL of the same ion or neutral atom such as [OIII]\,88\,$\mu$m/52\,$\mu$m, [NII]\,122\,$\mu$m/205\,$\mu$m, or [OI]\,145\,$\mu$m/63\,$\mu$m \citep[e.g.][]{spinoglio2015,herrera-camus2016}. Of these, we only have both [NII] lines, which allow to probe the density range 1\,cm$^{-3}$$<$$n_e$$<$$10^3$\,cm$^{-3}$. Our observed [NII]\,122\,$\mu$m/205\,$\mu$m luminosity ratio of 1.40$\pm$0.43 corresponds to an electron density $n_e=50\pm20$\,cm$^{-3}$ following the relation shown in for example Fig.~2 of \citet{herrera-camus2016}. This places SPT~0418-47 slightly above the average density found in nearby spirals and ultra-luminous galaxies from the ``Key Insights on Nearby Galaxies: A Far-Infrared Survey with {\it Herschel}'' (KINGFISH), ``Beyond the Peak'' (BtP), and GOALS samples \citep{herrera-camus2016,diaz-santos2017}. We are currently performing a survey of this [NII]\,122\,$\mu$m/205\,$\mu$m ratio in the SPT sample, which will be the first systematic survey in high-redshift galaxies, and refer to a forthcoming publication for a more in depth discussion (Cunningham et al., in prep.). 

It is important to keep in mind that the [NII] lines only trace the diffuse HII regions, but not the denser PDR. \citet{bothwell2017} modelled the observed ratio of four PDR-dominated lines (CO(2-1), CO(4-3), [CI]\,609\,$\mu$m, and [CII]\,158\,$\mu$m) with the {\tt 3D-PDR} code \citep{bisbas2012}, which also includes the effect of cosmic rays. They derive a consistent model with a PDR density of $n$$\sim$2.5$\times$10$^5$\,cm$^{-3}$, and a far-UV field strength $G0$$\sim$3.2$\times$10$^5$. This places SPT0418-47 at the high end of the observed PDR gas densities, while the $G0$ is the highest one in sample of 13 SPT DSFGs modelled by \citet{bothwell2017}, that is firmly in the region occupied by ULIRGs. As mentioned by \cite{bothwell2017}, the densities from the {\tt 3D-PDR} model are almost an order of magnitude higher than the ones derived using the PDR toolbox\footnote{http://dustem.astro.umd.edu/pdrt/} \citep{kaufman2006,pound2008}. The [OI]\,145\,$\mu$m line adds a powerful new diagnostic to the PDR modelling, as the [OI]\,145\,$\mu$m to [CII]\,158\,$\mu$m ratio is particularly sensitive to density, especially in the high $G0$ range. This ratio has the advantage of being mostly immune to extinction effects, but one has to keep in mind that it is also subject to abundance variations and HII contributions to [CII]\,158\,$\mu$m. From the PDR toolbox modelling of all five PDR-dominated lines, we derive a $G0$$\sim$$10^3$ and $n$$\sim$2$\times$10$^4$\,cm$^{-3}$. We defer a more detailed PDR modelling using {\tt 3D-PDR} to a future publication.

\subsection{AGN contributions}\label{subsection:AGNcontributions}

Optical line ratios are often used as an AGN diagnostic \citep[the classical ``BPT diagram''; ][]{baldwin1981}. In lensed DSFGs, this technique can unfortunately not be used as the optical spectra are dominated by the foreground lensing galaxy, while the background DSFG is almost completely dust obscured. At far-IR wavelengths, the situation is reversed: our sub-arcsecond 850$\mu$m ALMA imaging \citep{vieira2013} shows that the lens is undetected while the background DSFG is strongly detected. We can thus use the hot dust continuum shape and/or the mid- and far-IR FSL to look for AGN contributions.

Turning first to the dust continuum SED (Fig.~\ref{SED}), we find that it is consistent with the starburst template of \citet{magdis2012}, suggesting that an AGN is not required. To obtain a quantitative measure of the AGN fraction, we use the method described by \citet{spilker2018}, using the CIGALE SED modelling code \citep{noll2009,boquien2019} including the dust models of \citet{fritz2006}. We derive an AGN fraction $<$5\%. This is consitent with results in other DSFGs from the same SPT sample, where no significant AGN fractions were found \citep{ma2016,spilker2018,apostolovski2019}.


The alternative method we examine is to use the FSL. Several authors have used mid-IR and far-IR FSL to determine AGN contributions \citep[e.g.][]{fernandez-ontiveros2016,herrera-camus2018a}. These tracers are mostly based on the ratio of very high and lower ionization lines of the same species. Unfortunately, most of these species are in the mid-IR (e.g. [OIV]\,25.9\,$\mu$m, [NeIII]\,15.6\,$\mu$m, [NeII]\,12.8\,$\mu$m, [SIV]\,10.5\,$\mu$m, [SIII]\,18.7\,$\mu$m), and are inaccessible from the ground, even at high redshift. However, one ratio that does appear to separate AGN regions is the [OI]\,63\,$\mu$m / [CII]\,158\,$\mu$m line \citep[Fig. 12 of ][]{herrera-camus2018a}. While we did not observe the [OI]\,63\,$\mu$m line in SPT~0418-47, in Fig.~\ref{CII-OIII-OI-ratio}, we compare the [CII]\,158\,$\mu$m / [OI]\,145\,$\mu$m ratio in a variety of galaxies. The advantage of the 145\,$\mu$m line is that it is not affected by extinction and self-absorption. \citet{cormier2019} also showed that adding X-ray dominated regions to their cloudy models leads to a slight increase by a factor 1.2 in the [OI]\,145\,$\mu$m/63\,$\mu$m luminosity ratio (their Fig.~8). A similar effect is seen when increasing the cosmic ray flux. As AGN would produce both stronger X-rays and cosmic ray fluxes, the 145\,$\mu$m line is therefore at least as good an AGN tracer, if not a better one. In addition, AGN are expected to exite the higher [OI] level, that is increase the [OI]\,145\,$\mu$m/63\,$\mu$m ratio.

Figure~\ref{CII-OIII-OI-ratio} shows that the local AGN (blue symbols) and Seyferts (brown and a number of black symbols) cover the full range of [CII]\,158\,$\mu$m / [OI]\,145\,$\mu$m, but they are also the only ones with [CII]\,158\,$\mu$m / [OI]\,145\,$\mu$m $\lesssim$5. Interestingly, SPT~0418-47 falls among these lowest ratios, suggesting the possible presence of an AGN. However, we warn that this low [CII]\,158\,$\mu$m / [OI]\,145\,$\mu$m ratio may also be a direct consequence of the relatively high $T_{\rm dust}$=45$\pm$2\,K. Such higher dust and gas temperatures may be due to a higher Far-UV flux $G_0$, which will also increase [OI] faster than [CII] \citep[assuming the 145$\mu$m line behaves like the 63$\mu$m line,][]{malhotra2001}. Surprisingly, the AGN galaxies do not show higher [OIII]\,88\,$\mu$m / [OI]\,145\,$\mu$m ratios. This may be because we are comparing only galaxy-integrated line luminosities, and AGN emission is known to be very anisotropic due to an obscuring torus; the ionized gas is thus only observed in ionization cones and not throughout the galaxy. As such, the fact that SPT~0418-47 has a ratio within the region occupied by both AGN and normal galaxies does not provide any information on the presence of an AGN.
We conclude that while the SED seems to suggest a negligible AGN contribution in SPT~0418-47, the FSL ratio method is more inconclusive.

\subsection{Metallicity}\label{subsection:metallicity}

As argued by \citet{nagao2011} and \citet{pereira-santaella2017}, the ratio of [OIII]\,52\,$\mu$m + [OIII]\,88\,$\mu$m and [NIII]\,57\,$\mu$m is the most promising tracer of the gas metallicity using FSL. While these lines have now been observed with {\it Herschel} in a significant number of nearby objects, they are beyond the reach of ALMA for all but the very highest redshift objects known to date. More accessible alternatives are the ratios of [OIII]\,88\,$\mu$m with the [NII]\,122\,$\mu$m and/or [NII]\,205\,$\mu$m lines \citep{pereira-santaella2017,tadaki2019}. The main disadvantage of using the [NII] lines is that they have a much lower ionization potential of 14.53\,eV compared to 35.12\,eV for [OIII]\,88\,$\mu$m, creating a degeneracy between the ionization parameter $U$ and metallicity $Z$. An additional, though slightly less critical degeneracy exists with the gas electron density $n_e$. To solve this degeneracy between $U$ and $Z$, \citet{rigopoulou2018} use the ratio of the restframe 88/122\,$\mu$m continuum flux measurements. This assumes that the gas and dust are well mixed; the dust temperature $T_d$ will then increase with the ionizing flux, creating a correlation between the continuum flux ratio and $U$ \citep[see also][]{diaz-santos2017}. For PDR-dominated lines, a similar correlation between the Far-UV flux $G_0$ and the 60/100\,$\mu$m continuum flux ratio has indeed been observed in {\it ISO} data \citep[e.g.][]{malhotra2001,brauher2008}, suggesting the gas and dust temperatures are indeed correlated. 

\begin{figure}[ht]
\centering
\includegraphics[width=9cm]{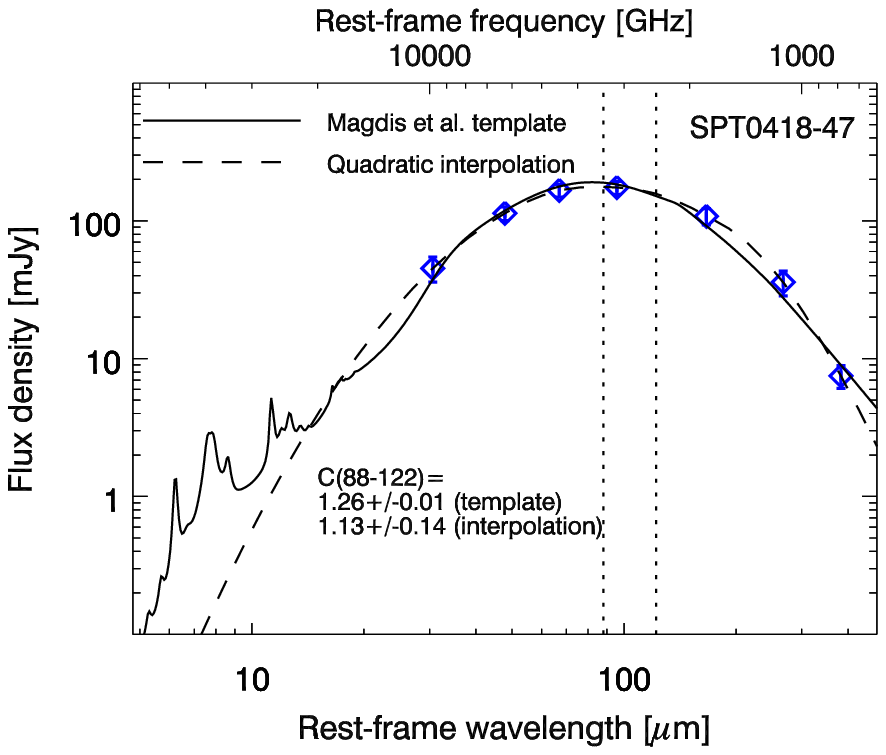} \\
\caption{Continuum Spectral Energy Distribution of SPT~0418-47 adjusted with the template of \citet{magdis2012} and a simple quadratic interpolation.}
\label{SED}%
\end{figure}

To allow a direct comparison with the method of \citet{rigopoulou2018}, we first derive the {\it rest frame} 88 and 122\,$\mu$m continuum flux ratio C(88-122) by fitting the continuum photometry of SPT~0418-47 (Table~\ref{photometry}) with the template of \citet{magdis2012} and a simple quadratic interpolation. Fig.~\ref{SED} shows that the results do not depend strongly on the adopted function, with C(88-122)=1.26$\pm$0.01 for the template and C(88-122)=1.13$\pm$0.14 for the quadratic interpolation. We conservatively adopt the latter, which yields $-$3.2$<$$\log U$$<$-2.0 using Fig.~4 of \citet{rigopoulou2018}. 

\begin{figure}[ht]
\centering
\includegraphics[width=9cm]{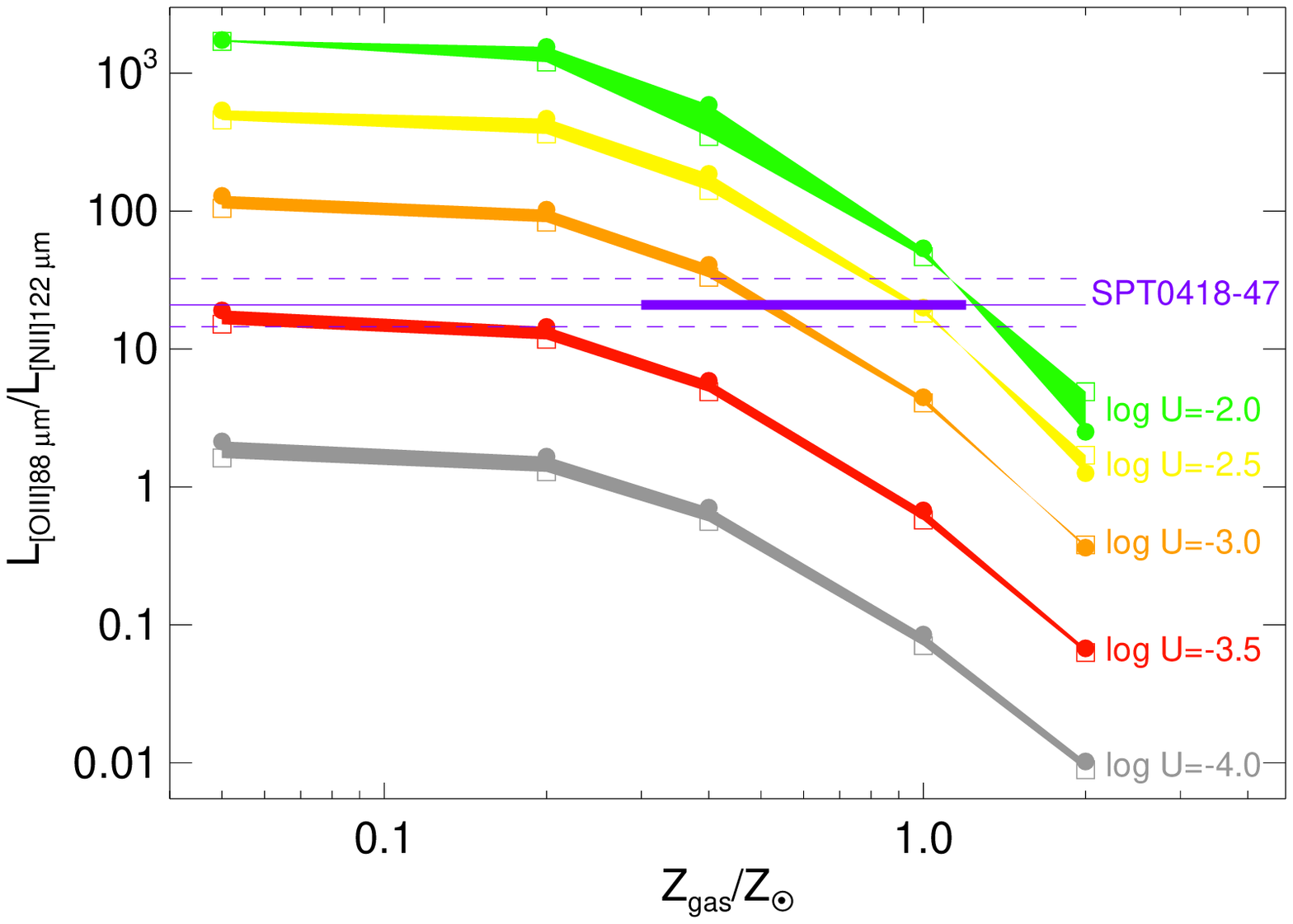} \\
\caption{[OIII]\,88\,$\mu$m over [NII]\,122\,$\mu$m line ratio as a function of metallicity. The colour shaded areas mark the range of ratios predicted by the CLOUDY models of \citet{pereira-santaella2017} with the same ionization parameter but different densities (circles $n$=10\,cm$^{-3}$ and squares $n$=10$^4$\,cm$^{-3}$). The colour coding (grey, red, orange, yellow, and green) indicates the ionization parameter (log U = -4.0, -3.5, -3.0, -2.5, and -2.0, respectively). The solid horizontal purple line marks the observed ratio in SPT~0418-47 with the uncertainty range delimited by dashed lines. The thick part of the horizontal line indicates the allowed range of log U as derived from the continuum SED (Fig.~\ref{SED}).}
\label{Z_OIIIoNII}%
\end{figure}

In Fig.~\ref{Z_OIIIoNII}, we reproduce the CLOUDY models of \citet{pereira-santaella2017}, where the coloured bands indicate different ionization parameters, and their spread different gas electron densities. The $\log U$ constraint limits the range of models to the orange, yellow and green shaded areas in Fig.~\ref{Z_OIIIoNII}. We note that there is very little dependence on the density, as the [OIII]\,88\,$\mu$m and [NII]\,122\,$\mu$m lines have very similar $n_{\rm crit}^{e^-}$=510 and 310\,cm$^{-3}$, respectively. The observed [OIII]\,88\,$\mu$m to [NII]\,122\,$\mu$m ratio thus suggests 0.3$<$$Z/Z_{\odot}$$<$1.3. This is consistent with the [CII]\,158\,$\mu$m / [NII]\,205\,$\mu$m ratio, which has also been used as a rough metallicity indicator \citep{nagao2012}. The ratio in SPT~0418-47 is lower than the one in ALESS~73.1 \citep{debreuck2014}, which also suggests a near solar metallicity.

Our new metallicity determination allows us to compare SPT~0418-47 to the mass-metallicity relationship studied at lower redshifts, and verify if this relation evolves over cosmic time. Although there is a rather large uncertainty, we derived a rather low stellar mass of $\sim$10$^{10}$M$_{\odot}$ (see \S\ref{subsection:nature}). The almost solar metallicity in SPT~0418-47 (corresponding to 8.18$<$12+log(O/H)$<$8.81) places it a bit higher on the mass-metallicity relation compared to other DSFGs at slightly lower redshifts \citep[Fig. 7 of ][]{rigopoulou2018}, near the track for $z$=2.2 from \citet{maiolino2008} and \citet{mannucci2010}. 

%

\section{Conclusions}
We have detected five FSL in SPT~0418-47, including the second detection of [OI]\,145\,$\mu$m at $z$$>$1. Most of the analytic power of the FSL comes from line ratios. We draw the following conclusions from our study:
Firstly, the ratio of [OIII]\,88\,$\mu$m to [OI]\,145\,$\mu$m suggests that the emission from the ISM in SPT~0418-47 is dominated by PDR rather than HII regions. The [CII]\,158\,$\mu$m line is almost 90\% dominated by PDR.
Secondly, the ratios of [CII]\,158\,$\mu$m to [OI]\,145\,$\mu$m and of [NII]\,205\,$\mu$m to [NII]\,122\,$\mu$m suggest that both the PDR and HII regions have high gas densities.
Thirdly, using a new diagram of [CII]\,158\,$\mu$m to [OI]\,145\,$\mu$m against [OIII]\,88\,$\mu$m to [OI]\,145\,$\mu$m, we find that SPT~0418-47 lies in a region occupied by nearby AGN in low redshift samples. While this suggests that SPT~0418-47 may also harbour an AGN, we warn that the observed line ratios may also be a direct consequence of the relatively warm $T_d$=45.3$\pm$2.3\,K created by an extremely powerful starburst.
Fourthly, following the method of \citet{rigopoulou2018}, we derive an ionization parameter $-$3.2$<$$\log U$$<$-2.0 from the dust continuum flux ratios. Combined with the observed [OIII]\,88\,$\mu$m to [NII]\,122\,$\mu$m line ratio, we derive a gas metallicity of 0.3$<$$Z/Z_{\odot}$$<$1.3 in SPT~0418-47.

Taken together, these results imply that SPT~0418-47 has a dense, roughly solar metallicty ISM that is dominated by PDR. The presence of an AGN cannot be excluded, but it is equally possible that the extreme strength of the starburst is heating both the dust and gas in the entire galaxy to higher than average values.

The high gravitational amplification ($\mu$=32.7$\pm$2.7) of SPT~0418-47 has allowed us to detect even the weaker [OI]\,145\,$\mu$m line with APEX. This line has been often neglected at high redshift, probably because it was also difficult to study in nearby samples before the advent of {\it Herschel}. It is a powerful alternative to its brighter counterpart at 63\,$\mu$m, being at a significantly easier frequency to observe from the ground at $z$$>$1.17, rather than $z$$>$4 for the 63\,$\mu$m line. Moreover, the 145\,$\mu$m line is not subject to extinction and self-absorption affects, which may reduce the observed flux of the 63\,$\mu$m line. We expect the [OI]\,145\,$\mu$m line to become a powerful new tracer of PDR in high-$z$ galaxies.

\begin{acknowledgements}
We thank the anonymous referee for useful suggestions, Myha Vuong for stimulating discussions on fine structure lines, and the APEX staff for their excellent support in obtaining the data. This publication is based on data acquired with the Atacama Pathfinder Experiment (APEX) under programme IDs 098.A-0513 (ESO), 0101.F-9522, and 093.F-9512 (both Max Planck). APEX is a collaboration between the Max-Planck-Institut fur Radioastronomie, the European Southern Observatory, and the Onsala Space Observatory.
This paper makes use of the following ALMA data: ADS/JAO.ALMA\#2016.1.00133.T and ADS/JAO.ALMA\#2018.1.01670.S. ALMA is a partnership   of ESO (representing its member states), NSF (USA) and NINS (Japan), together with NRC (Canada), MOST and ASIAA (Taiwan), and KASI (Republic of Korea), in cooperation with the Republic of Chile. The Joint ALMA Observatory is operated by ESO, AUI/NRAO and NAOJ.
The SPT is supported by the NSF through grant PLR-1248097, with partial support through PHY-1125897, the Kavli Foundation and the Gordon and Betty Moore Foundation grant GBMF 947.
The National Radio Astronomy Observatory is a facility of the National Science Foundation operated under cooperative agreement by Associated Universities, Inc.
\end{acknowledgements}

%
%

\end{document}